\documentclass[aps,prb,twocolumn,groupaddress,floatfix,showpacs,showkeys]{revtex4}
\usepackage{graphicx}
\usepackage{dcolumn}
\usepackage{bm}
\usepackage{amsmath}
\usepackage{multirow}
\usepackage[sort&compress]{natbib}
\usepackage{hyperref}
\usepackage{hypernat}
\usepackage{xcolor}

\hypersetup{
  citebordercolor  = {0 0 0},
  linkbordercolor = {0 .5 .5},
  bookmarks=false,
  colorlinks,
  linkcolor=blue,
  citecolor=blue,
  pdftitle={CT satellites in x-ray spectra of transition metal oxides v 9.2},
  pdfauthor={Klevak},
  urlcolor=blue
}

\newcolumntype{f}[1]{D{.}{.}{#1}}

\begin{document}

\title{Charge transfer satellites in x-ray spectra of transition metal oxides}

\date{\today}
\author{E. Klevak, J. J. Kas and J. J. Rehr}
\affiliation{Department of Physics, University of Washington, Seattle, WA 98195}

\begin{abstract} 
%Strongly correlated materials such as TM oxides often exhibit
%strong satellites in their x-ray photoemission (XPS) and x-ray absorption
%spectra (XAS) due to localized charge-transfer (CT) excitations
%that accompany the sudden creation of a core hole.
Strongly correlated materials such as transition metal oxides (TMOs)
often exhibit
large satellites in their x-ray photoemission (XPS) and x-ray absorption
spectra (XAS). These satellites arise from localized charge-transfer (CT)
excitations that accompany the sudden creation of a core hole.
%These excitations
%are treated here in a two-step approach with a localized system 
%embedded in a condensed system coupled to a photoelectron.
Here we use a two-step approach to treat such excitations
in a localized system embedded in a condensed system and coupled to a
photoelectron.
The total XAS is then given by a convolution of a spectral function
representing the localized excitations and the XAS of the extended
system.  The local system is modeled roughly in terms of a simple
three-level model, leading to a double-pole approximation
for the spectral function that represents dynamically weighted
contributions from the dominant neutral and charge-transfer excitations.
This method is implemented using a resolvent approach,
with potentials,
radial wave-functions and  matrix elements from the real-space Green's
function code \textsc{feff}, and parameters fitted to
XPS experiments.  Representative calculations for several 
TMOs are found to be in reasonable
agreement with experiment.

\end{abstract}

\date{\today}

\maketitle

\section{\label{sec:intro}Introduction}
Shake-up excitations in x-ray absorption (XAS) and x-ray 
photoemission spectra (XPS)
have long been of interest.\citep{Aberg,Sawatzky1,Sawatzky2}
These excitations are generated by the intrinsic response of a system to a
suddenly created core hole, and are reflected in satellite peaks
in the spectra.  Examples in XAS range from edge-singularities
in metals,\citep{ND} to many-body amplitude
factors in x-ray absorption fine structure.\citep{CRFEFF}
Recently such satellites have also been found to explain the
extrinsic and intrinsic losses and interference effects 
in XPS experiments.\citep{Guzzo,martensson} 
However, these effects are relatively small in weakly-correlated
materials, where the dominant excitations are plasmons. In those cases the
satellite amplitudes are of order 10\% of the main peak
and become negligible near absorption thresholds
in the adiabatic limit.\citep{Campbell} 
Consequently, broadened single-particle theories with a core hole 
can be good approximations.\citep{FEFF,DSCF,DSCFRev,Stobe}
In contrast, dramatic satellites comparable in strength to the main peak
are typically observed in the spectra of correlated
materials such as TMOs and high-temperature superconductors.
These satellites are often attributed to localized charge-transfer (CT)
excitations.\citep{deGrootbook}
In such cases the one-particle
approximation fails dramatically in the near-edge region.
Several approaches with various degrees of sophistication have been
introduced to address this behavior.
For example, the ``charge-transfer multiplet approach" treats
strong correlations locally, with solid-state effects modeled by
crystal-field parameters.\citep{deGrootarticle1}
Configuration interaction techniques have also been
applied to small clusters,\citep{Ikeno} but
these methods are computationally intensive.
%Dynamical mean field approaches have also been employed starting from
%the Anderson impurity model.[DMFT]\citep{DMFT}

Our goal in this work is to develop a simple yet practical, semi-quantitative
approach
to model both local correlations and solid-state effects to explain these
excitations. Our approach is based on a simplified two-step model
with a localized system embedded
in a solid, and coupled to a photoelectron. The approach incorporates
both localized charge-transfer excitations and long-ranged plasmon
excitations. In particular, our method combines the model
of localized excitations introduced by Lee, Gunnarsson and
Hedin (LGH),\citep{LHedin} with the treatment of solid-state effects and
other inelastic losses as in the real-space Green's function approach
used in the \textsc{feff9} XAS code.\citep{FEFF} As a justification for
this separation we note that the localized and extended excitations are
spatially and energetically decoupled.
% Parameters for the local model are derived from experiment.
Our main result is an expression for the XAS of charge-transfer systems
as a convolution of an effective spectral function $A_L(\omega,\omega')$
that contains the localized CT excitations, and an approximation for
the XAS of extended systems $\tilde\mu(\omega)$ that builds in
long-range, extrinsic inelastic losses
\begin{equation}
\label{mufirst}
\mu(\omega)=\int d\omega'\, A_L(\omega,\omega')\, \tilde\mu(\omega-\omega')
\equiv A_L * \tilde \mu.
\end{equation}

As discussed by Kas et al.,\citep{kas2007many}
%the effective
%XAS that takes into account inelastic losses
$\tilde\mu(\omega)$ is 
related to the quasi-particle XAS 
by an analogous convolution $\tilde \mu = A_Q * \mu$.
%which includes extrinsic losses
%from long-range (e.g., plasmon) excitations.
At low energies compared
to the plasma frequency, plasmon satellites become negligible and
$\tilde \mu(\omega) \approx \mu(\omega)$, i.e.
%i.e., the quasi-particle XAS spectrum $\mu(\omega)$
the spectra calculated in the presence of a core hole.  Within the simplest
three-level model for the localized system, the CT spectral function
$A_L$ has two energy-dependent peaks separated by a characteristic
charge-transfer energy splitting $\delta E$ which is typically a few eV.
%The nature of the cross-over from the adiabatic to sudden transition also
%differs from the case with plasmons. 
Our result in Eq.\ (\ref{mufirst}) is similar to the formulation of
Calandra et al.,\citep{Calandra} where the spectral function is taken to
be the XPS spectra $A_L(\omega,\omega')=\sigma(\omega-\omega')$. 
In contrast the present approach makes use of an explicit model for
the localized system and also approximates dynamic effects, such as the
crossover from
the adiabatic to the sudden limit.  We have applied this method
systematically to a number of 3d TMOs, and obtain results in
reasonable agreement
with experiment and other calculations.\citep{Calandra,wu_quadrupolar_2004}

\section{Theory}
\label{sec:theory}
\subsection{LOCAL MODEL}
\label{sec:local_model}
\begin{figure}[t]
\includegraphics[width=8.6cm,clip]{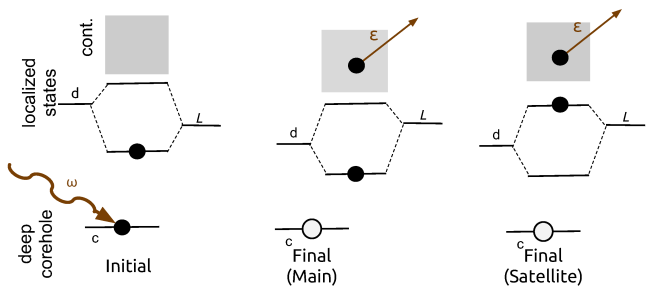}
\caption{\label{fig:model_fig}
Schematic representation of the three-level model (core state and two localized 
states): the labels are
d - transition metal
3d level; L - ligand valence state, cont - continuum states;
 $\varepsilon$ - photoelectron kinetic energy.}
\end{figure}

Our model for the localized system is adapted from the three-level 
tight-binding model of Lee, Gunnarsson and Hedin (referred to here as
LGH), \cite{LHedin} which is only briefly summarized here.
For clarity we adopt similar notation and some key formulae are reproduced
in the Appendix; we refer to original paper \citep{LHedin} for additional
details.
%This model
%(referred to here as LGH)
%can be generalized straightforwardly
%to a finite number of localized levels,
%we consider only
%three levels.
The LGH model can be extended to a more realistic description,
for example, using the Haydock recursion
scheme\citep{haydock_surface_1973} applied to a tight-binding 
Hamiltonian, and keeping only the leading iterations. 
Nevertheless, the simplified LGH model captures the essential physics of the
charge-transfer process.  As illustrated in 
Fig.\ \ref{fig:model_fig}, the levels include a strongly localized $d$ level,
a less localized ligand level $L$, and a deep core level $c$.
Physically, this local model represents a system in which upon photoecxitation,
a localized level $d$ is pulled below the ligand level $L$ due to
the Coulomb interaction with the core hole. As a result, there is
a finite probability that the electron originally in the $L$ state is
transferred to the $d$ state. This process corresponds to the lowest energy
main peak (``shake-down") in the photoemission process and
strongly screens the core hole. There is also a finite probability
that the electron will remain in level $L$, corresponding to the satellite
peak and a less screened core hole. The effective spectral function is
determined from the relative probabilities of these two processes.
\\\indent The Hamiltonian of the full system is separated as
\begin{equation}
\label{fullhamiltonian}
\mathcal{H} = \mathcal{H}_{0} + T + V + \Delta,
\end{equation}
where ${\cal H}_0$ is the Hamiltonian of the local system, $T$
is the kinetic energy of the photoelectron, $V$ is the
Coulomb interaction between the photoelectron and the local system,
and $\Delta$ is the coupling to the x-ray field. In detail
\begin{eqnarray}
\label{model}
{\cal H}_0 &=& \sum_i \varepsilon_i n_i + \sum_l U_{l} n_{c} n_{l} 
+ t(c^{\dagger}_{d} c_{L} + hc), \nonumber \\
V &=&  \sum_{{\bf kk}'}\left [\sum_i n_i V^i_{{\bf kk}'} - V^c_{{\bf k k}'} \right] c_{\bf k}^{\dagger} c_{\bf k'}, \nonumber \\
\Delta &=& \sum_{\bf k} M_{c{\bf k}}\, [c_{\bf k}^{\dagger}c_c + hc\,].
\end{eqnarray}
Here $i=(c,d,L)$, $l=(d,L)$,
$\varepsilon_{d},\, \varepsilon_{L}$, and $\varepsilon_{c}$ are,
respectively, the bare energies of the $d$, $L$ and $c$ levels,
$U_{d}$ and $U_{L}$
represent, respectively, the Coulomb interaction between the
core hole and the $d$ and $L$ levels, and $t$ is a hopping matrix 
element, which we approximate by just one value, $k$ and $k^{'}$ are wave
vectors of continuum states. 
Implicit in this model is the constraint
$n_d+n_L=1$, so that ${\cal H}_0$ can be simplified in terms of a single
Hubbard-like
parameter $U=U_d-U_L$. Similarly, $n_c=0$ when the core hole is present
and only the difference in the potentials
$V_{sc}({\bf r}) =V_L({\bf r}) -V_d({\bf r})$ is needed in Eq.\ (\ref{Vsc_pot}) to
represent the change in potential when the electron hops from the ligand
level $L$ to the localized
$d$ level. The scattering matrix elements $V_{kk^{'}}$ are given by
$\langle \vec k |V_{sc}({\bf r}) |\vec k'\rangle$, and
$U =\int d^3 r\, \rho_c({\bf r}) V({\bf r}) \approx V(0)$,
since the core charge $\rho_c$ is highly localized.
Throughout this paper we will use Hartree atomic units $e=\hbar=m=1$
unless otherwise specified.

\begin{figure}[h]
\includegraphics[width=8.6cm,clip]{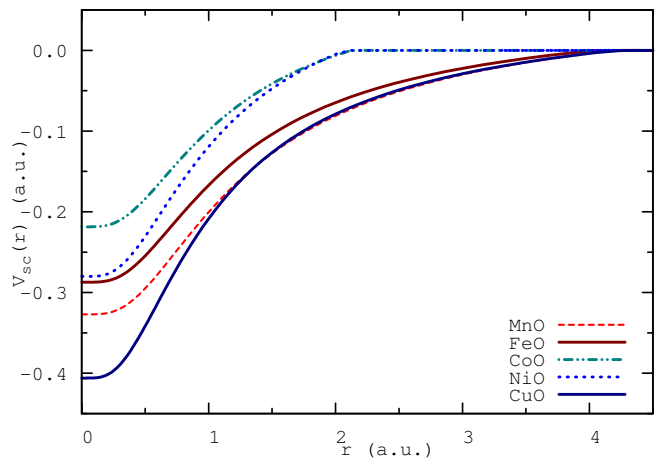}
\caption{\label{fig:Vsc}
Photoelectron scattering potentials $V_{sc}(r)$ see Eq.\ \ref{Vsc_pot} for MnO, FeO, CoO, 
NiO and CuO}
\end{figure}

\begin{figure}[b]
\includegraphics[width=8.6cm,clip]{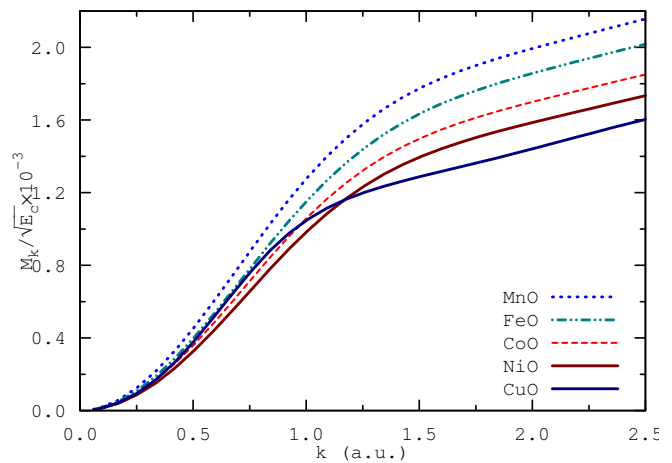}
\caption{\label{fig:Mk}
%\textcolor{red}{[THIS FIGURE IS NEVER REFERENCED IN THE TEXT]}
Dipole matrix elements $M_{ck}$ in Eq.\  (\ref{Mk})
for MnO, FeO, CoO, NiO and CuO, calculated from \textsc{feff}.}
\end{figure}

\begin{figure}[t]
 		\includegraphics[width=8.6cm,clip]{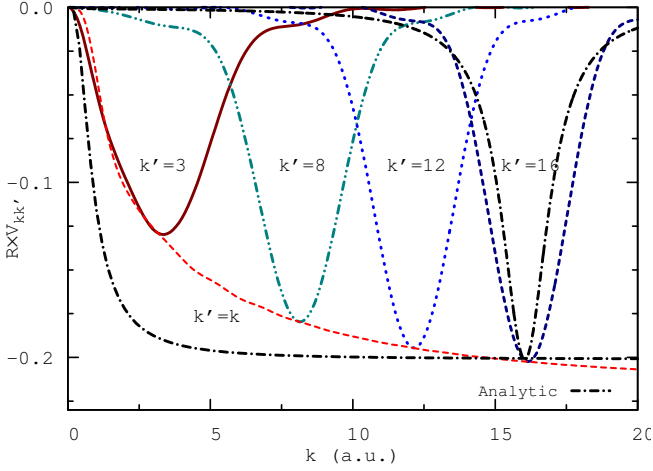}
\caption{\label{fig:Vkk}
	Matrix elements $V_{kk^{'}}$ for
	CoO. For comparison, the LGH analytical form [Ref. \onlinecite{LHedin}]
  of $V_{kk'}$ for $k=k^{'}$ and $k^{'}=16$ au	is also shown,
	calculated with parameters: $\tilde{V}=-0.79$
  a.u., $R_{sd} = 1.01$ au and $R_s = 3.97$ au.}
\end{figure}

The excited states of this local model $|\Psi_{sk}^f\rangle$, with
$s=1$ and 2, can be calculated exactly within a two-particle basis
$|\psi_s\rangle|\psi_k\rangle$, using the resolvent approach of LGH 
\begin{equation}
 \label{resolv_final}
 |\Psi_{sk}^{f} \rangle = \left[1+\frac{1}{E - \mathcal{H}_0 - T - V - i \eta}V\right]|\psi_s\rangle|\psi_k\rangle,
\end{equation}
where $|\psi_s \rangle$ correspond to the eigenstates of $\mathcal{H}_0$ 
[see Eq.\ (\ref{finalstates})] with a full core-hole $n_c=0$.
Following LGH we also represent the local potential by 
\begin{equation}
 \label{Vsc_pot}
 V_{sc}({\bf r}) =
 \begin{cases}
 [-V_{3d}(r)+\frac{1}{R_0}]/\varepsilon_0 & r < R_{0} \\ 0  & r > R_{0}.
 \end{cases}
\end{equation}
Here $V_{3d}(r)$ is the potential of the $d$ level calculated using
3d wave functions from \textsc{feff9},\citep{FEFF} and the $1/R_0$ term
crudely represents the potential of the ligand charge shell.
The constant $\varepsilon_0$ is chosen so that $U =
-V_{sc}(0)$.
Since $V_{3d}(0)\gg1/R_{0}$, $\varepsilon_0 \approx V_{3d}(0)/U$.
The parameters used in our sample calculations are given in Table \ref{table1}, 
with details on how they are obtained given in Sec.\ ~\ref{sec:CT_Parameters}.
\renewcommand{\arraystretch}{1.5}
The scattering potentials $V_{sc}(r)$ are shown in Fig.\ \ref{fig:Vsc}.
These parameters are calculated using \textsc{feff} wave functions and
values of $R_0$ from Table \ref{table1}
which represent distances between absorber and ligand atoms.
The scattering potential is set to zero beyond $R_0$.
\begin{table}[h]
\begin{center}
\caption{\label{table1}
Parameters used in the local model:
$U,t$ and $\delta E$ for MnO, FeO, CoO, NiO and CuO are fit to XAS
experiments (Ref. \onlinecite{Calandra,wu_quadrupolar_2004}, see Sec.\ \ref{sec:CT_Parameters}),
%for reference values for CuCl$_{2}$ can be found in LGH \cite{LHedin}. 
$R_0$ is obtained by averaging the distance between 
absorber and ligand atoms, $\varepsilon_0, \delta E, \phi$ and $r_{00}$
were calculated using Eq.\ (\ref{energy}), (\ref{tanphi}) and (\ref{r00}).}
\begin{ruledtabular}
%\begin{tabular}{  lddddddf{6}r }
\begin{tabular}{ lcccccccc }
      & $U$(eV) & $t$(eV) & $R_{0}$(a.u.) & \vline & $\varepsilon_0$ & $\delta E$(eV) & $\phi=\theta$ & $r_{00}$\\
\hline
  %      CuCl$_{2}$ & 10.58 & 1.84 & 4.71 & & 1.96 & 6.45 & 0.35 & 1.43 \\
  %             U      t     R0       eps0   dE    phi    r00
	MnO        & 11.0 & 1.6 & 2.23 & & 3.16 & 6.3 & 0.26 & 3.0\\
	FeO        &  9.7 & 1.9 & 2.14 & & 3.83 & 6.1 & 0.33 & 1.7\\
	CoO        &  8.9 & 1.5 & 2.13 & & 4.38 & 5.3 & 0.29 & 2.3\\
	NiO        & 10.6 & 2.1 & 2.08 & & 3.65 & 6.8 & 0.34 & 1.5\\
	CuO        & 13.0 & 1.5 & 2.23 & & 3.03 & 7.2 & 0.21 & 5.4\\
\end{tabular}
\end{ruledtabular}
\end{center}
\end{table}
The matrix elements $V_{kk'}$ of the scattering potential
are presented in Fig.\ \ref{fig:Vkk} and 
compared with the analytic form discussed in LGH. Note that the numerical
calculations are in good agreement with the analytical form except at low
energies.
%We also compare matrix element with the fitted analytic form for the matrix element of the 
%scattered potential $V_{kk'}$ discussed in LGH
%(see Fig.\ \ref{fig:Vkk}), and find good agreement except at low energies for
%CuCl$_{2}$ and CoO.
The matrix elements $V_{kk'}$ for NiO and CoO are similar, due
to the similarity of their scattering potentials $V_{sc}(r)$ (see
Fig.\ \ref{fig:Vsc}).
%To match analytical form of scattering matrix elements \ref{fig:second}
% to the exact \ref{fig:first} we use the parameter $L$ to represent the size
%of the radial box which is large on an atomic scale.
The radial transition matrix elements (see Fig.\ \ref{fig:Mk})
are given by
%$M_k = \langle c |r |k\rangle$ are 
\begin{equation}
 \label{Mk}
 M_{ck} \propto \sqrt{(\varepsilon_k - \varepsilon_c)} \int \mathrm{d}{r}\, r^2
\, \psi_{c}(r)\,r\,u_{lk}(r),
\end{equation}
where the radial wave functions $u_{lk}(r)$  are
\begin{equation}
\label{wavefnct_sol}	
 u_{lk}(r)=
 \begin{cases}
\tilde  u_{lk}(r) & r < r_{mt}   \\
   \sqrt{2/L}\, [h_l(kr)e^{i\delta_{lk}}+c.c.]  & r > r_{mt},
\end{cases}
\end{equation}
and are calculated from subroutines in 
\textsc{feff9}\citep{FEFF}.
Here, $\tilde u_{lk}(r)$ corresponds to
the regular solution for angular momentum
$l$ at the origin, matched to solutions beyond
$r_{mt}$ in terms of spherical Hankel functions $h_l(kr)$, and
$\delta_{lk}$ is the partial wave phase shift. Boundary conditions within
a sphere of large radius $L=40$ a.u. 
and an exponential grid with $0.01$ a.u. step are used
to obtain smooth results for the matrix elements.
%Representative results for CoO are given in Fig.\ \ref{fig:Mk}.

\begin{figure}[t]
\includegraphics[width=8.6cm,clip]{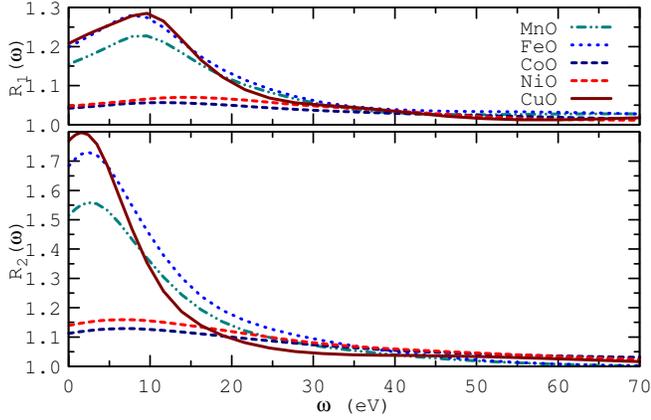}
\caption{
\label{fig:R_i_all}
Ratios between full photocurrent Eq.\ (\ref{current}) and photocurrent in the 
sudden limit Eq.\ (\ref{e:photocurrent_sudden}) for the main $R_1(\omega)$ (top) and
satellite $R_2(\omega)$ (bottom) peaks Eq.\ (\ref{R_s}) 
for MnO, FeO, CoO, NiO and CuO.  Note the significant dynamic
variation at low energies while the adiabatic limit is reached
around $30$ eV.
}
\end{figure}

\begin{figure}[h!]
 \includegraphics[width=8.6cm,clip]{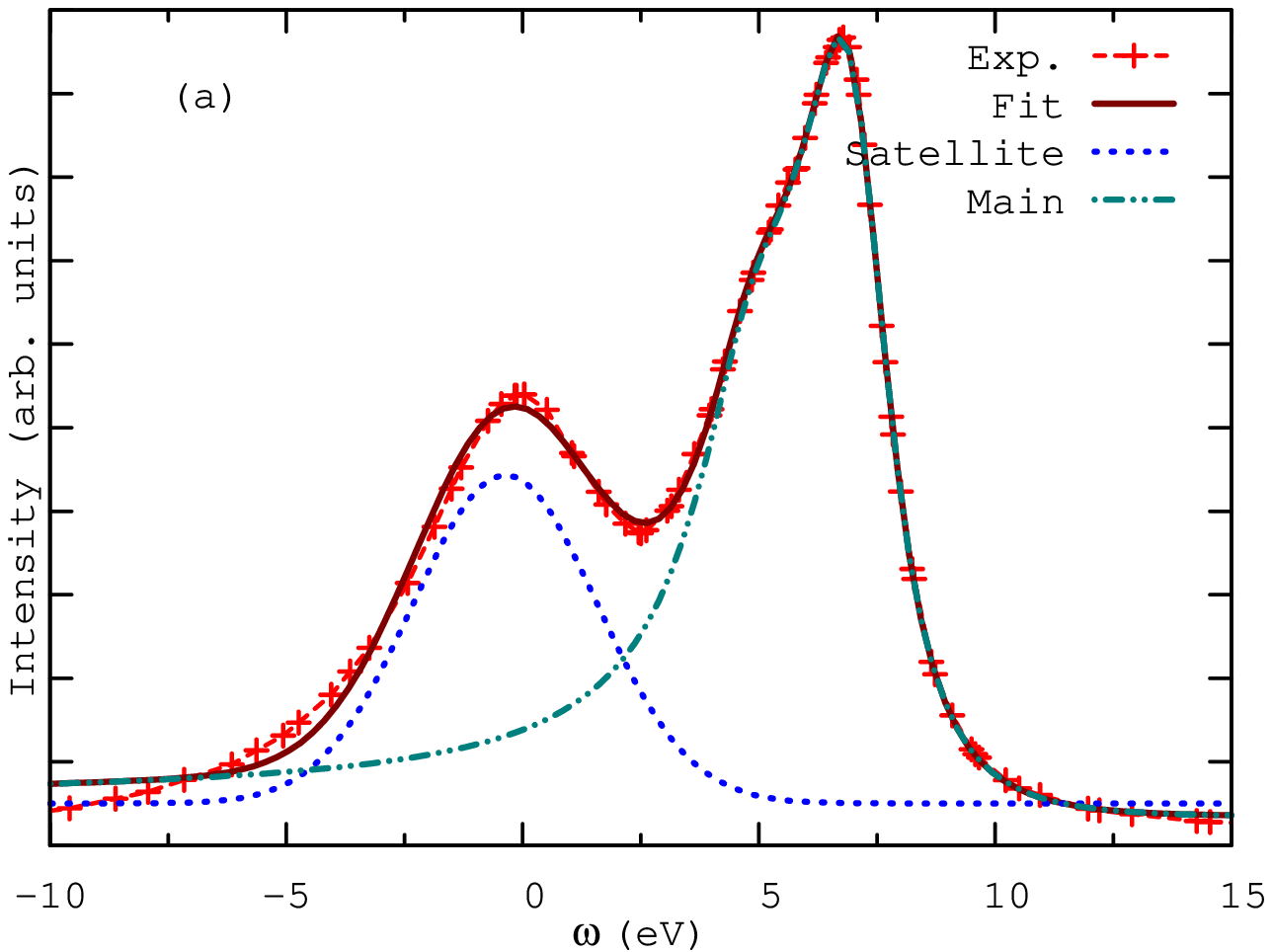}
 \includegraphics[width=8.6cm,clip]{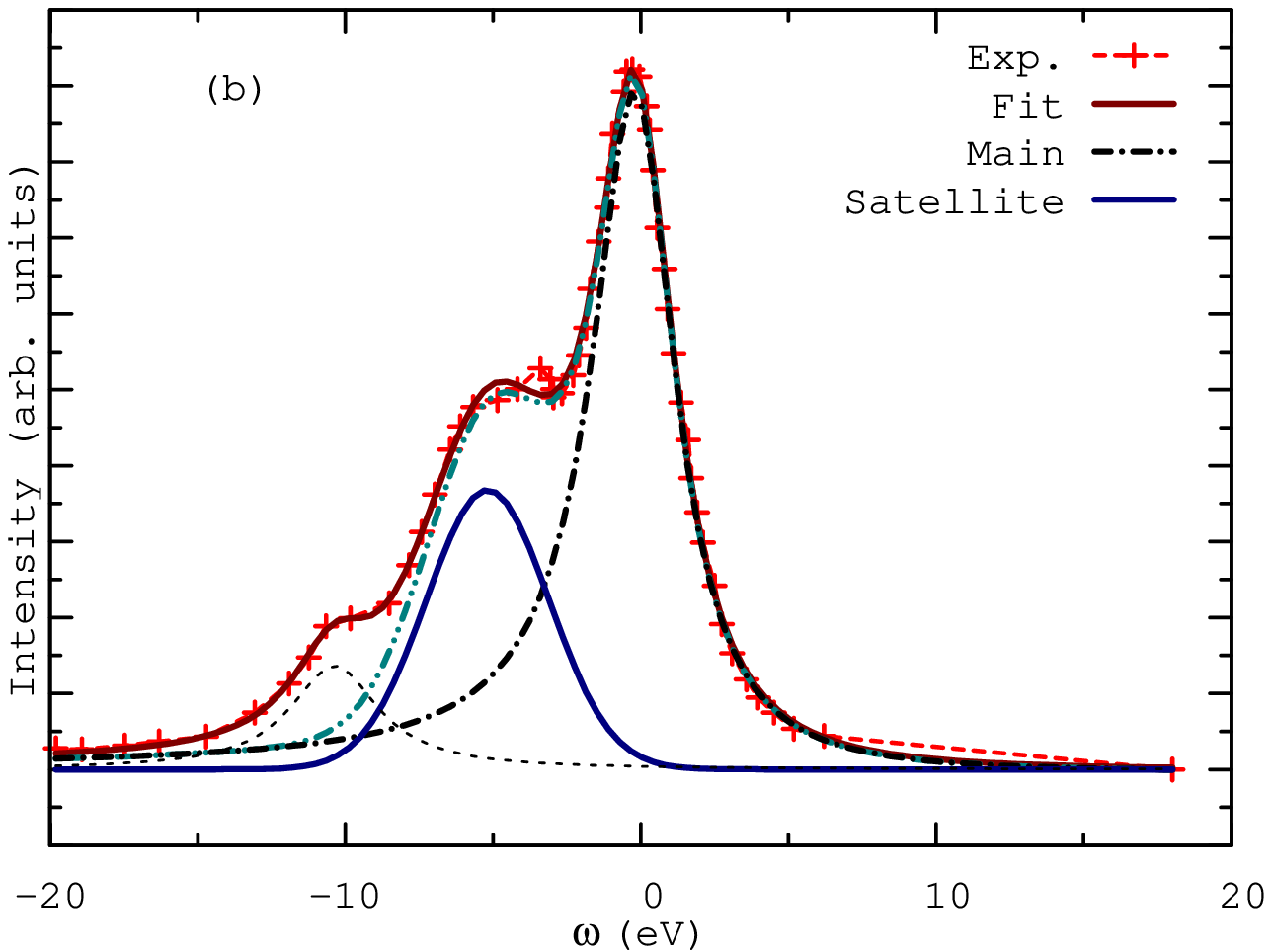}
 \caption{
(a) Recent XPS experiment [Ref. \onlinecite{Calandra}] of NiO 1s edge,
with subtracted plasmon background. (b) XPS spectra [Ref. \onlinecite{Fulvio}] for CoO.
}
 \label{fig:XPS_exp}
\end{figure}

%\begin{figure}[h]
%\includegraphics[width=8.6cm,clip]{./data/CoO/CoO_xps.pdf}
%\caption{\label{fig:CoO_XPS}
%XPS spectra\citep{Calandra} for CoO.
%}
%\end{figure}

\subsection{CT PARAMETERS}
\label{sec:CT_Parameters}
Due to the simplicity of our model, there is no simple
correspondence to the tight-binding parameters of a more realistic system.
%\textcolor{red}{Due to the simplicity of our model, we do not expect a
%direct correspondence between its parameters and those in a more
%realistic one.}
Nevertheless, the charge-transfer parameters of the three level system can
be chosen to fit the main ($s=1$) and satellite ($s=2$)
peaks in XPS experiments.
The energy difference between these two peaks is defined as
$\delta E$, while $w_1/w_2$ refers to the ratio of intensity of the
main to the satellite peak and is given by
%is the ratio of the areas under the peaks;
Eq.\ (\ref{energy}) and (\ref{ws}).
For NiO we fit these quantities to the 
1s XPS edge data (Fig.\ \ref{fig:XPS_exp} a)
of Calandra et al.\citep{Calandra}
Due to band splitting, the main peak is bimodal and asymmetric. 
Thus we simply fit its strength to two complex Lorentzians,
while the satellite peak was fit with a single Gaussian. 
The plasmon peaks at about $-25$ eV are ignored,
since they are implicitly included in $\tilde\mu$ and are only important
at high energies.  Solving
Eq.\ (\ref{energy}) and (\ref{ws})
%then
yields estimates for $U$ and $t$.
For CoO we used 3s XPS edge data,\citep{Fulvio} since 1s results are not
available.
As in the case of NiO, we also subtracted the peak near
$-12$ eV. We fit the main peak with a single complex Lorentzian while
the satellite was fit with a single Gaussian. In the case of MnO, FeO and CuO
we used 3s XPS experimental data \citep{galakhov_x-ray_1999} as 1s are not
available. For MnO and FeO we fit the main and satellite peaks with
complex Lorentzians, following the same procedure for estimating
parameters for the LGH model described above.
We also tried to estimate the hopping parameter $t$ from
the width of the projected $d$-density of states of the absorber 
as obtained from the \textsc{ldos} module in \textsc{feff}, but
these results only agree qualitatively with the fits given
in Table \ref{table1}.

\section{Results}
\label{sec:results}
Assuming isotropic XPS and summing over all directions,
the XAS is simply related to the XPS photocurrent 
%by a factor of $1/k_i$,
\begin{equation}
 \label{absspectra}
 \mu(\omega) = \sum_{sk} J_{k}^{s}(\omega)  
\propto\sum_s\frac{1}{k_s} | M(s,k_s)|^2 ,
\end{equation}
where $M(s,k) \equiv \langle\Psi_{sk}^{f}|\Delta|\Psi_0\rangle$
%\begin{eqnarray}
% \label{dipole_matrix}
%  M(s,k) &\equiv& \langle\Psi^{sk}_{f}|\Delta|\Psi_0\rangle =
%    \langle\psi_k|\langle\psi_s|\bigg[1+  \\ \nonumber
%       &+& \left. V\frac{1}{\epsilon_k + E_s -
% \mathcal{H}_0 -T -V +i\eta}\right]\Delta|\Psi_0\rangle,
%\end{eqnarray}
is obtained from the resolvent formula in Eq.\ (\ref{resolv_final})), where
$k_s = \sqrt{2(\omega+E_0-E_s)}$,  and $E_s$ are obtained from eigenvalues
of the model Hamiltonian from Eq.\ (\ref{energy}).
To simplify the discussion of the dynamical effects,
we introduce the ratio $R_s(\omega)$
between the calculated photocurrents at a given photon energy,
and the photocurrent $J^0_i(\omega)$ in the sudden approximation,
\begin{equation}
\label{R_s}
	R_s(\omega) = \sum_k J^s_{k}(\omega)/ J^0_{s}(\omega).
\end{equation}
where the sudden-limit is
\begin{equation}
\label{e:photocurrent_sudden}
 \begin{split}
  J^{0}_{s}(\omega)&= \frac{1}{k_s}\sum_k |M_{ck} w_s|^2
  \delta(\omega-\epsilon_k+E_0-E_s)\\
  & = w_s^2\, \tilde\mu(\omega-E_s),
 \end{split}
\end{equation}
and $J^s_k(\omega)$ is given my Eq.\ (\ref{current}).
Here $\tilde\mu$ implicitly includes the effects
of long-range inelastic losses in the XAS, $M_{ck}$
is given by Eq.\ (\ref{Mk}),
and the weights $w_s$ are given by Eq.\ (\ref{ws}).
Thus the total XAS can be written as
\begin{equation}
 \label{spectral_fn}
 \begin{split}
 \mu(\omega)& = \sum_{sk}{|M(s,k)|^2}\delta(\omega-\epsilon_k+E_0-E_s)\\
 &= \sum_{s} J_{s}^{0}(\omega) R_{s}(\omega).
 \end{split}
\end{equation}
Fig.\ \ref{fig:R_i_all} shows $R_s(\omega)$ for a number of TMOs. Note that there is a 
significant
``overshoot" at low energies, while above about 30 eV, $R_s(\omega)$ tends
to the sudden limit.
This behavior arises from the interplay between intrinsic and extrinsic
effects represented by the first and second terms in 
Eq.\ (\ref{resolv_final}).  The overshoot is a fairly small correction
to the adiabatic limit, since the interaction
between the scattered electron and the core hole is relatively small as a
result of screening of the core hole by the charge transfer process.
\begin{figure}[t!]
    \includegraphics[width=8.6cm,clip]{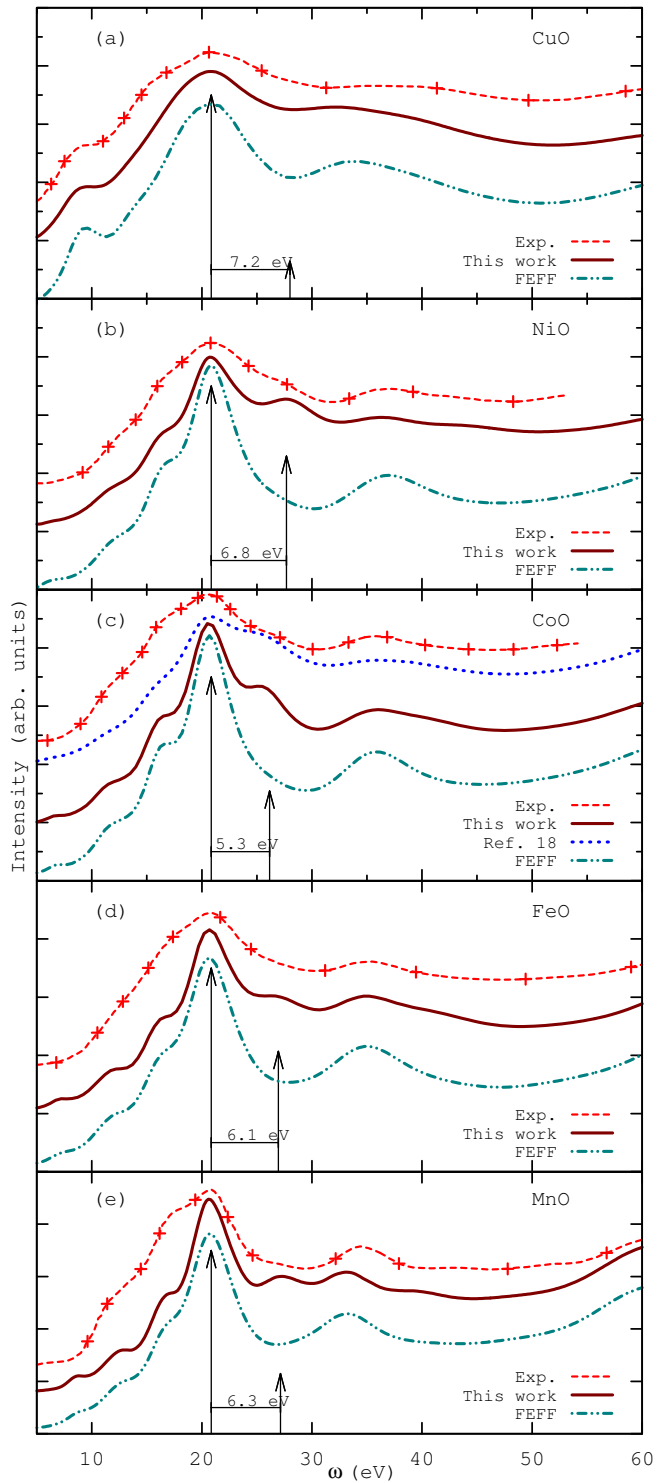}
\caption{
Comparison of the experimental XAS
[Ref. \onlinecite{Calandra,wu_quadrupolar_2004}], convolution using three level model (LGH), to
single particle (\textsc{feff}) calculation 
for (a) MnO, (b) FeO, (c) CoO, (d) NiO and (e) CuO. And Calandra approach [Ref. 
\onlinecite{Calandra}] for CoO. Main and satellite peak position and intensities
are shown with arrows. Intensity ratio is equal to $r_{00}$, see Table \ref{table1}}.
\label{fig:XAS_all_1}
\end{figure}
In our calculations,
$\tilde \mu(\omega-E_s)$ is the total XAS spectrum calculated
from \textsc{feff9}.\citep{FEFF}
Thus one finally obtains the convolution formula of Eq.\ (\ref{mufirst}) with
the spectral function $A_L(\omega,\omega^{'})$
%where
given by
\begin{equation}
 \label{conv_fun}
  A_L(\omega,\omega') =\frac{1}{D}
\sum_s {w_s^2 R_s(\omega)\delta(\omega-\omega' -\omega_s)},
\end{equation}
where the normalization constant $D =  \sum w_s^2 R_s(\omega)$.
Calculations of $R_s(\omega)$ were performed using the resolvent
formula (see Eq.\ (\ref{resolv_final})) with
a Hamiltonian matrix with indices $s,k$ and 80 $k$-points
of dimension $160 \times 160$.
Calculations of the XAS for NiO were carried out using
both our two step model and, for comparison, a convolution with a
multiple Lorentzian fit to the XPS as in Calandra et al.\citep{Calandra}
Results for CoO
are presented in Fig.\ \ref{fig:XAS_all_1} (c) and show that
the two methods are numerically similar, with
small differences arising from differences in broadening between two experiments. 
To account for differences
in broadening ($\Delta\Gamma$) of the main and satellite peaks in CoO, which can be 
seen in XPS experimental data (see Fig. \ref{fig:XPS_exp}), broadening
of the satellite peak was carried out with value $\Delta\Gamma=0.4$ eV. 
Results for CuO and NiO are presented in Fig.\ \ref{fig:XAS_all_1} (a, b) using values of 
$\Delta\Gamma$ $0.6$ and $1$ eV respectively.
%convolution indeed broadening in former method is peak dependent.
Calculations of XAS for FeO and MnO are presented in Fig.\ \ref{fig:XAS_all_1} (d, e),
values of $\Delta\Gamma$ are $1.46$ and $1.1$ eV respectively.
The effect of CT satellites are clearly seen in the Fig.\ \ref{fig:XAS_all_1}.
In particular the satellite peak transfers oscillator strength from the main
peak and fills in missing spectral weight in one particle calculations
above the main peak. On the other hand, the treatment here is only
semi-quantitative as the satellite peaks seen in the experimental XAS
are more broadened than can be represented by a two-peaked spectral function.
%Peaks arising due to the CT are labeled as ``Satellite". 
For MnO, the parameters for the LGH model were fit to the XPS experiments shown in
Fig.\ \ref{fig:XPS_exp}, and are given in Table \ref{table1}.
%Due to
%To match broadenings in experimental XAS data, broadenings of
%1.57 eV and 1.43 eV was used in \textsc{feff} for NiO and CoO, respectively.
%Similarly to account for the finite life time of a core hole in case of MnO, broadening
%of 1.5 eV was used to produce the final result, additionally broadening of 2 eV was used
%for satellite peak to account for many possible transitions from d level. In case of
%FeO broadening of 2 eV was used due to the life time of a core hole and 2eV for satellite 
%peak due to the same reasoning. 

\begin{figure}[t!]
    \includegraphics[width=8.6cm,clip]{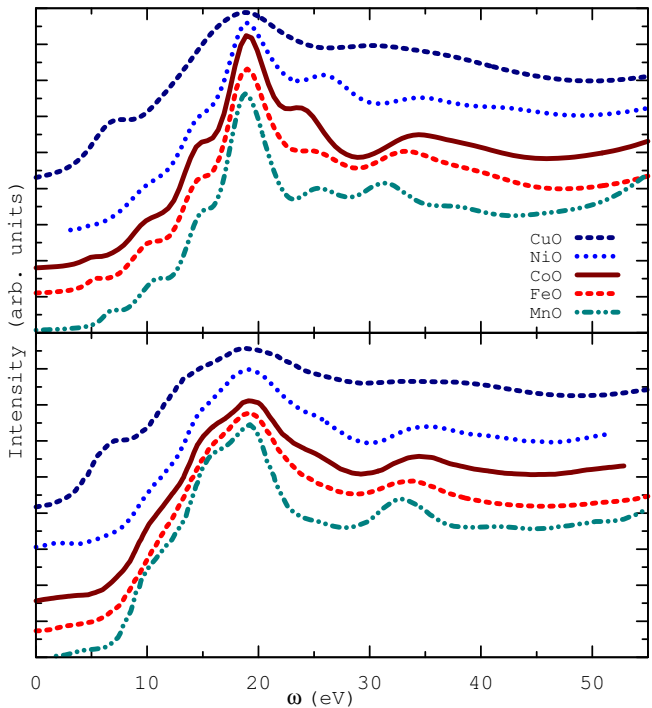}
\caption{ \label{fig:XAS_all_3}
Comparison of spectra calculated using three level system (LGH) (top)
and experimental [Ref. \onlinecite{wu_quadrupolar_2004,Calandra}] 
XAS data (bottom) for a number of TMOs.}
\end{figure}

\section{\label{sec:conclusions}Conclusions}
We have developed a simplified, semi-empirical model of
the effects of charge-transfer excitations in XAS, thus
%We have
extending the formulation of Lee, Gunnarsson, and Hedin.\citep{LHedin}
%in which 
The spectra are modeled by a localized three-state system
coupled to a photoelectron, and implemented using the
\textsc{feff9} real-space
Green's function code to include solid state and extrinsic losses.
The final spectrum is a convolution of a single-particle XAS calculated
using \textsc{feff}, with a frequency-dependent spectral function
consisting of two delta functions that represents the localized 
charge-transfer excitations.
%which
%This model takes into account extrinsic as well as intrinsic effects. 
%We have compared results of XAS calculations for the metal
%K edges of NiO and CoO to experimental data and find good
%agreement.
We find fairly good agreement between our results and the XAS 
at the metal K edges for number of TMOs (see Fig.\ (\ref{fig:XAS_all_3})).
In these spectra, the presence of charge transfer satellites are 
clearly seen by comparing the total and single-particle XAS, where such
peaks are missing in the latter.  The convolution of the single-particle
spectra from \textsc{feff}  with the CT spectral function $A_L(\omega)$
reproduces fairly well the peaks
%\textcolor{red}{creates}
%in the spectrum
at higher energies with an energy splitting $\delta E$. However,
the CT satellite peaks in the model spectra are sharper, which is likely
an artifact of the two-delta function model for the spectral function.
%This is probably due to the fact that,
%This suggests that the extrinsic effects are small; indeed they only become
%significant for photoelectron energies large compared to typical plasmon
%excitations of order 20 eV. In cases when extrinsic effects are large
%we may need to extend the approximations used here, e.g., 
%by treating the total spectrum as a sum of two
%spectra corresponding to well-screened and poorly-screened
%core-photoelectron interactions. 
From Fig.\ (\ref{fig:XAS_all_3}) one can see that there is a noticeable
discrepancy between
experimental and calculated spectra in the pre-edge region: in some cases there
are missing peaks, while in others the intensities are smaller. These
differences might be due to the spherical muffin-tin\citep{PhysRev.51.846} 
approximation of the scattering potential, as full potential single 
particle calculations\citep{Calandra} for CoO and NiO have greater intensities
in the pre-edge region. However the main goal of this
paper was to approximate the CT satellite peaks.
%separated by characteristic energy shift $\delta E$
Further work is needed to obtain \textit{ab initio}
values of the parameters used in the model. In principle these could be
found using constrained DFT or constrained RPA methods.

\begin{acknowledgments}
We thank S. Baroni, C. Brouder, K. Jorissen, F. Manghi, L. Reining,  S. Story,
and F. D. Vila for comments and suggestions. This work is supported in
part by the DOE Grant DE-FG03-97ER45623 (JJR) and was facilitated by
the DOE Computational Materials Science Network.
\end{acknowledgments}

\appendix
\section{\label{sec:app_lmodel}LOCAL MODEL}
Here we briefly summarize the details of the sudden approximation
for the local model, closely following the methodology and notation 
of LGH.\citep{LHedin}
The initial state $|\Psi_0\rangle$ is the ground state of $\mathcal{H}_{0}$
with $n_c=1$  
\begin{equation}
\label{initialstate}
 |\Psi^0\rangle = -\sin(\theta)|\psi_c\rangle|\psi_a\rangle +
	\cos(\theta)|\psi_c\rangle|\psi_b\rangle  .
% |\Psi_{sk}^{f} \rangle &=& |\psi_k\rangle | \psi_s\rangle \\
\end{equation}
The final states $|\Psi_{sk}^f\rangle$ are given by
Eq.\ (\ref{resolv_final}),
where $|\psi_1\rangle$ and $|\psi_2\rangle$ 
are the eigenstates of $\mathcal{H}_{0}$ with $n_c=0$, and
can be parameterized conveniently in terms of mixing angles $\theta$ and
$\phi$
\begin{eqnarray}
\label{finalstates}
|\psi_s\rangle &=&  
\begin{cases} 
	\cos(\phi)|\psi_{a}\rangle -
           \sin(\phi)|\psi_{b}\rangle,\mbox{ }s=1,  \\ 
	\sin(\phi)|\psi_{a}\rangle +
           \cos(\phi)|\psi_{b}\rangle,\mbox{ }s=2, \\
	\end{cases}
\end{eqnarray}
where 
\begin{eqnarray}
\label{tanphi}
 \tan(2\theta) &=& 2t/(\varepsilon_a - \varepsilon_b + U), \nonumber \\ 
 \tan(2\phi) &=& 2t/(\varepsilon_b - \varepsilon_a),
\end{eqnarray}
and $a$ and $b$ correspond to the metal d- and oxygen p-levels.
%\begin{eqnarray}
% \label{initialstate_1}
% |\psi_{1}\rangle &=& \cos(\phi)|\psi_{a}\rangle -
%\sin(\phi)|\psi_{b}\rangle,\\
% |\psi_{2}\rangle &=& \sin(\phi)|\psi_{a}\rangle +
%\cos(\phi)|\psi_{b}\rangle,
%\end{eqnarray}
%\begin{equation}
% \label{initialstate_1}
% |\psi_{s}\rangle = 
%			\begin{cases} 
%				\mbox{ } \cos(\phi)|\psi_{a}\rangle -\sin(\phi)|\psi_{b}\rangle,\mbox{ } s = 1 \\ 
%				\mbox{ } \sin(\phi)|\psi_{a}\rangle +\cos(\phi)|\psi_{b}\rangle,\mbox{ } s = 2.
%			\end{cases}
%\end{equation}
The photocurrent is then calculated using\citep{synch}
\begin{equation}
 \label{current}
 J^{s}_{k}(\omega) = |\langle \Psi_{sk}^{f}
|\Delta| \Psi^{0}\rangle |^2 \delta(\omega - \varepsilon_k + E_0 - E_s).
\end{equation}
%where $|\Psi_{sk}^{f} \rangle$ is the
%where the final state of the two electron system
%with one electron in the superposition
%state and the second in a scattered state $|\psi_k\rangle$.
The spectrum of the model Hamiltonian is characterized by the parameters
\begin{equation}
 \label{energy}
  \begin{split}
  &E_s = \frac{1}{2}(\varepsilon_a + \varepsilon_b) \mp \delta E / 2,\\
  &\delta E= \sqrt{(\varepsilon_a - \varepsilon_b)^2 + 4 t^2}.
  \end{split}
\end{equation}
We consider only the symmetric case
with $\varepsilon_a = \varepsilon - U/2$ and $\varepsilon_b = \varepsilon$. 
%So after the transition
%we have shake-down. Since before the transition a level is above the b level, while after the transition b level is above a level. 
%To obtain the absorption spectra we sum the kinetic energy distribution of
%the photoelectron
%\begin{eqnarray}
% \label{absspectra2}
% \mu(\omega) &=& \sum_{sk} J_{k}^{s}(\omega)  \propto
% \sum_s \frac{1}{k_s} | M(s,k_s)|^2,\\ 
%  k_s &=& \sqrt{2(\omega + E_0 - E_s)}.\nonumber
%\end{eqnarray}
%where the $1/k_s$ the Jacobian from functional dependence of delta function.
%Here $M_k$ is the dipole matrix element given by Eq.\ (\ref{Mk}) and 
%given by (\ref{Mk}) and for $\omega_s$ we have.
The weights of main and satellite levels are then
\begin{equation}
 \label{ws}
 w_s = 
	\begin{cases} 
		-\sin(\phi+\theta),\mbox{ } s = 1 \\ 
		\mbox{ }\mbox{ }\cos(\phi+\theta),\mbox{ } s = 2. 
	\end{cases}
\end{equation}
In the sudden\citep{LHedin} limit there is no interaction between the
photoelectron and the electron on the outer level, so that
the ratio of the main to the satellite peak intensities is
\begin{equation}
 \label{r00}
 r_{00} = \lim_{\omega \to \infty}
\frac{\sum_k J^2_k(\omega)}{\sum_k J^1_k(\omega)} = \cot^{2}(\phi + \theta).
\end{equation}
%where $r_{00}$ represents
%Taking into account the analytical form of dipole matrix elements in the
%sudden\citep{LHedin} limit one obtains
%[{\bf EGOR: do you need the following equation?? Where is it used?}]
%\begin{equation}
% \label{rw_suden}
% r(\omega) = r_{00} \left[\frac{\tilde{\omega}}{\tilde{\omega}+\delta E}\right]^{\frac{3}{2}}
%\left[\frac{1+(\tilde{\omega} + \delta E)/ \tilde{E_{d}}}{1+\tilde{\omega}/\tilde{E_d}}\right]^{2},
%\end{equation}
%where $\tilde{E_d}$ is characteristic energy of the analytic dipole matrix element $M_k$, 
%$\tilde{\omega} = \omega -\omega_{th}= \omega -(E_2 - E_1) $ and $\delta E$ is given in (\ref{energy}).
%\begin{figure}[h]
%\includegraphics[width=8.6cm,clip]{./data/resolv.ps}
%\caption{\label{fig:resolv_fig}
%$r(\omega)$ at $R=1$, sudden approximation at different characteristic energies $\tilde{E_d}$
%}
%\end{figure}
%\section{Exact treatment}
%Finally, from Eq.\ (\ref{resolv_final}) the transition matrix element
%$\langle\Psi_{sk}^{f}|\Delta|\Psi_0\rangle$ 
%for the two state model is 
%\begin{eqnarray}
% \label{dipole_matrix}
%M(s,k) &\equiv& \langle\Psi_{sk}^{f}|\Delta|\Psi_0\rangle =
%    \langle\psi_k|\langle\psi_s|\bigg[1+  \\ \nonumber
%       &+& \left. V\frac{1}{\epsilon_k + E_s -
% \mathcal{H}_0 -T -V +i\eta}\right]\Delta|\Psi_0\rangle.
%\end{eqnarray}

%\bibliography{CTsatellites}
\bibliographystyle{apsrev}

\end{document}